\documentclass[nofootinbib,prd,twocolumn,showpacs,showkeys,preprintnumbers]{revtex4-1}
\usepackage{hyperref,amssymb,amsmath,mathrsfs,bm,graphicx}
\begin{document}
\title {Deconstructing Frame Dragging}
\author{L. Herrera}
\email{lherrera@usal.es}
\affiliation{Instituto Universitario de F\'isica
Fundamental y Matem\'aticas, Universidad de Salamanca, Salamanca 37007, Spain}
\date{\today}

\begin{abstract} The vorticity of world lines of observers associated to the rotation of a massive body was  reported by Lense and Thirring more than a century ago. In their example  the frame dragging effect  induced by  the vorticity,   is directly  (explicitly) related to the rotation of the source. However in   many other  cases it is not so, and the origin of vorticity  remains obscure and difficult to identify.  Accordingly, in order to unravel this issue, and looking for the ultimate origin of vorticity associated to frame dragging,  we  analyze in this manuscript very different scenarios where frame dragging effect is present. Specifically we consider general vacuum stationary spacetimes, general electro--vacuum  spacetimes, radiating electro--vacuum spacetimes and Bondi--Sachs radiating spacetimes. We  identify the physical quantities present  in all these cases, which   determine  the vorticity  and may legitimately be considered as responsible for  the  frame dragging. Doing so we provide a comprehensive  physical picture of frame dragging. Some observational consequences of our results are discussed.
\end{abstract}
\date{\today}
\pacs{04.20.-q; 04.20.Cv; 04.30.Nk}
\keywords{Frame dragging; Super--energy; Gravitational radiation}  
\maketitle

\section{Introduction}
  
The dragging of inertial frames produced by self--gravitating sources, whose existence  has been recently  established by observations \cite{1},  is one of the most remarkable  effect predicted by the general theory of relativity (GR) (see \cite{1n, 2n}).

The term frame dragging usually refers to the influence of a rotating massive body on a gyroscope by producing  vorticity in the congruence of world lines  of observers outside the rotating object. Although the appropriateness of the term ``frame dragging'' has been questioned by Rindler \cite{2b}, it nevertheless has been used regularly in the literature till nowadays, and accordingly we shall adopt such term here (see also \cite{3n, 4n}).

The basic concept for the understanding of this effect is that of vorticity  of a congruence, which  describes the rotation of a  gyroscope attached to the congruence,  with respect to reference particles.  

Two different effects may be detected by means of gyroscopes. One of this (Fokker--de Sitter effect) refers to the precession of a gyroscope following a closed orbit around a spherically symmetric mass distribution. It has been verified with a great degree of accuracy by observing the rotation of the earth--moon system  around the sun \cite{2c}, but  this is not the frame dragging effect we are interested in here.  The other effect, the one we are concerned with in this work,  is the Lense--Thirring--Schiff precession, which  refers to the appearance of vorticity in the congruence of world lines of observers in the gravitational field of a massive rotating ball. It was reported  by the first time by Lense and Thirring \cite{3}, and is usually referred to as Lense-Thirring effect ( some authors suggest that it should be named instead, Einstein-Thirring-Lense effect, see \cite{5n, 6n, 7n}).
This result led Schiff \cite{4} to propose the use of gyroscopes to measure such an effect. Since then this idea has been developed extensively (see \cite{2b, 4b, 5, 7, 8, 10, 10a, 11, 12, 13} and references cited therein).

However, although the origin of vorticity may be easily  identified in the Lense--Thirring metric, as due to the rotation of a massive object, it is not always explicitly  related to rotation of massive objects. In fact, in any vacuum stationary space time (besides the Lense--Thirring metric) we can detect a frame dragging  effect, without specifying  an explicit  link to the rotation of a massive body \cite{13b}.

The situation is still more striking for the electro--vacuum space times. The point is that the quantity responsible for the rotational (relativistic) multipole moments in these spacetimes,  is  affected by the   mass rotations (angular momentum), as well as by the electromagnetic field, i.e.  it contains contributions from both (angular momentum and electromagnetic field). This explains why  such a quantity does not necessarily vanish in the case when the angular momentum of the source is zero but electromagnetic fields are present. 

The first known example of  this kind of situation was brought out  by Bonnor \cite{14}. Thus, analyzing the gravitational field of a magnetic dipole plus an electric charge, he showed that the corresponding spacetime is stationary    and a frame dragging effect appears. As a matter of fact  all stationary electro--vacuum solutions exhibit frame dragging \cite{15}, even though  in some  cases   the angular momentum of the source is zero. In this latter case the rotational relativistic multipole moments  and thereby the vorticity,   are generated by the electromagnetic field. Furthermore, as we shall see, electrodynamic radiation also produces vorticity.

Finally, it is worth recalling that vorticity is present  in gravitationally radiating space--times. The influence of gravitational radiation on a gyroscope through the vorticity associated with the emission of gravitational radiation was put forward for the first time in \cite{16}, and has been discussed in detail since then in \cite{17, 18, 19, 20, 21, 22, 23, 24}. In this case too, the explicit relationship between the vorticity and the emission of gravitational radiation was established without resorting to the rotation of the source itself.

Although in many of the scenarios described above a rotating object is not explicitly identified as the source of vorticity, the fact remains that  at purely intuitive level, one always associates the vorticity of a congruence of world lines, under any circumstance, to the rotation of ``something''. 

The purpose of this work is twofold: on the one hand we shall identify the physical concept (the ``something'') behind all cases where frame dragging is observed, whether or not the angular momentum of the source vanishes. On the other hand we would like to emphasize the possible observational consequences of our results.

As we shall see below, in all possible cases, the appearing  vorticity is accounted for  by the  existence of a flow of superenergy on the plane orthogonal to the vorticity vector, plus (in the case of electro--vacuum spacetimes) a  flow of electromagnetic energy on the same plane.

Since superenergy plays a fundamental role in our approach, we shall start by  providing a brief introduction of this concept in the next section.

\section{Superenergy and super--Poynting vector}

The concept of energy is a fundamental tool in all branches of  physics, allowing to approach and solve a vast number of  problems under a variety of circumstances.  This explains the fact that since the early times of  GR  many researchers have tried by means of very different approaches to present a convincing definition of  gravitational energy, in terms  of an invariant local quantity
All these attempts, as is well known, have failed. The reason for this failure is easy to understand.

Indeed, as  we  know, in classical field theory energy is a quantity
defined in terms of potentials and their first derivatives. On the other hand however, we also know that in GR it is impossible to construct a tensor
expressed only through the metric tensor (the potentials) and their first derivatives (in accordance with  the
equivalence principle). Therefore, a local description of
gravitational energy in terms of true invariants (tensors of any
rank) is  not possible within the context of the theory.

Thus, the following alternatives remain:
\begin{itemize}
\item  To define energy in terms of a non--local quantity.
\item  To resort to  pseudo--tensors.
\item  To introduce a succedaneous definition of energy.
\end{itemize}

One example of the last of the above alternatives is superenergy,   which may be defined
either from the Bel  or from the  Bel--Robinson tensor  \cite{16b, 17b, 18b}
(they both coincide in vacuum), and has been shown to be very useful
when it comes to explaining  a number of  phenomena in the context of
GR.

Both, the Bel and the Bel--Robinson tensors, are obtained by invoking  the ``structural'' analogy between GR and  Maxwell theory of electromagnetism.  More specifically, exploiting  the analogy between the  Riemann  tensor ($R_{\alpha \beta \gamma \delta}$) and the  Maxwell  tensor ($F_{\mu \nu}$), Bel introduced a four--index  tensor  defined in terms of  the Riemann tensor in a way which is a reminiscence of the definition of the energy--momentum tensor of electromagnetism   in terms of the Maxwell tensor. This is the Bel tensor.

The Bel--Robinson tensor is defined as the Bel tensor, but with the Riemann tensor replaced by the Weyl tensor ($C_{\alpha \beta \gamma \delta}$)  (see \cite{19b} for a comprehensive account and more recent references on this issue).

Let us now introduce the electric and magnetic parts of the Riemann and the Weyl tensors  as,
\begin{equation}
E_{\alpha \beta}=C(R)_{\alpha \gamma \beta \delta}u^{\gamma}u^{\delta}, \label{elec}
\end{equation}

\begin{equation}
H_{\alpha \beta}=C(R)^\ast _{\alpha \gamma \beta \delta}u^{\gamma}u^{\delta},
\label{mag}
\end{equation}

 where $C(R)_{\alpha \gamma \beta \delta}$ is the  Weyl (Riemann) tensor, the four--vector $u^{\gamma}$ in vacuum is the tangent vector to the  world--lines of observers, and  $C(R)^\ast _{\alpha \gamma \beta \delta}$ is the dual of the Weyl(Riemann) tensor.
 
A third tensor may be defined from the double dual of the Riemann tensor as 
\begin{equation}
  X_{\alpha \beta}={^{\ast}}R^\ast _{\alpha \gamma \beta \delta}u^{\gamma}u^{\delta},
  \label{herma}
  \end{equation}
which in the case of the Weyl tensor $ X_{\alpha \beta}$ coincide with the electric part of the Weyl tensor (up to a sign).

Next, from the analogy with electromagnetism the super--energy and the super--Poynting vector are defined by 
\begin{eqnarray}
&&U(R)=\frac{1}{2}(X_{\alpha \beta} X^{\alpha \beta}+E_{\alpha \beta} E^{\alpha \beta})+H_{\alpha \beta} H^{\alpha \beta},\nonumber \\
&&U(C)=E_{\alpha \beta} E^{\alpha \beta}+H_{\alpha \beta} H^{\alpha \beta},
\label{super1}
\end{eqnarray}
\begin{eqnarray}
&&P(R)_\alpha=\eta_{\alpha \beta \gamma \delta}(E^\beta_\epsilon H^{\gamma \epsilon}-X^{\beta}_{\epsilon} H^{\epsilon \gamma})u^\delta,\nonumber \\
&&P(C)_\alpha=2\eta_{\alpha \beta \gamma \delta}E^\beta_\epsilon H^{\gamma \epsilon}u^\delta.
\label{supepo}
\end{eqnarray}
where  $R(C)$ denotes whether the quantity is defined with Riemann (Weyl) tensor, and $\eta_{\alpha \beta \gamma \delta}$ is the Levi--Civita tensor.

In the next sections we shall bring out the role played by the above introduced variables in the study of the frame dragging effect.
\section{Frame dragging in vacuum stationary spacetimes}
As we mentioned in the Introduction, the first case of frame dragging analyzed in the literature was the Lense--Thirring effect. For didactical reasons we shall start by considering first this  case and from there on, we shall consider examples of increasing complexity. Thus, afterward we shall consider the Kerr metric, an approximation of which is the Lense--Thirring spacetime, and finally we shall consider the general vacuum stationary spacetime case.

\subsection{The Lense--Thirring precession}
The Lense--Thirring effect  is based on an approximate solution to the Einstein equations which reads  \cite{3}
\begin{widetext}
\begin{eqnarray}
ds^2=-\left(1-\frac{2m}{r}\right)dt^2+\left(1+\frac{2m}{r}\right)\left(dr^2+r^2d\theta^2+r^2\sin^2\theta d\phi^2\right)
+\frac{4J\sin^2\theta}{r}d\phi dt. \label{LT1}
\end{eqnarray}
\end{widetext}
It describes the gravitational field outside a spinning sphere of
constant density, up to first order in $m/r$ and
$J/r^2$, with $m$ and $J$ denoting the mass and the angular momentum
respectively.

Up to that  order, it coincides with  the Kerr
metric, by identifying 
\begin{equation}
ma=-J \label{ltk}
\end{equation}
where $a$ is the Kerr parameter \cite{20b}.

Next, the congruence of the world--lines of observers at rest in the
frame of (\ref{LT1}) is described by the timelike vector $u^{\alpha}$ whose components are
\begin{equation}
u^{\alpha}=\left(\frac{1}{\sqrt{1-\frac{2m}{r}}},0,0,0\right),
\label{LTu}
\end{equation}

from the above expression  the vorticity vector, defined as usual by
\begin{equation}
\omega^\alpha=\frac{1}{2}\eta^{\alpha\eta\iota\lambda}u_\eta
u_{\iota,\lambda},\label{vv}
\end{equation}
has, up to order $a/r$ and $m/r$, the following non--null components
\begin{equation}
\omega^r=\frac{2ma\cos\theta}{r^3}, \label{w1}
\end{equation}
\begin{equation}
\omega^\theta=\frac{ma\sin\theta}{r^4}, \label{w2}
\end{equation}
or 
\begin{equation}
\Omega=(\omega^\alpha
\omega_{\alpha})^{1/2}=\frac{ma}{r^3}\sqrt{1+3\cos^2\theta},
\label{LT2}
\end{equation}

which at  $\theta=\frac{\pi}{2}$ reads
\begin{equation}
\Omega=\frac{ma}{r^3} \label{LT3}.
\end{equation}

The above expression embodies the essence of the Lense--Thirring effect. It describe the vorticity of the world lines of observers, produced by the rotation  ($J$)  of the source.  Such vorticity, as correctly guessed by Shiff \cite{4}, could be detected by a gyroscope attached to the world lines of our observer. 

Even though  in this case the vorticity is explicitly related to the rotation of the spinning object which sources the gravitational field,  the fact that this link in many other cases is not so explicitly  established leads us to the question: what  is (are) the physical mechanism(s) which explains the appearance of vorticity  in the world lines of the observer? As we shall see in the next sections the answer to this question may be given in terms of a flow of  superenergy plus  (in the case of electro--vacuum spacetimes) a flow of electromagnetic energy.

So,  in order to approach to this conclusion, let us calculate    the leading term of the super--Poynting gravitational vector at the equator. Using (\ref{supepo}) and (\ref{LT1}) we obtain  for the only non--vanishing component (remember  that in vacuum both expressions for the super--Poynting vector coincide)

\begin{equation}
P^\phi \approx 9\frac{m^2}{r^2} \frac{a}{r}\frac{1}{r^5},\label{LT3bis}
\end{equation}
It describes a flux of super--energy on the plane orthogonal to the vorticity vector. On the other hand it follows at once from (\ref{LT3bis}) that

$P^\phi=0  \Leftrightarrow a=0
\Leftrightarrow\omega^{\alpha} = 0$.

From the comments above, a hint about the link between superenergy and vorticity (frame dragging) begins to appear. In order to delve deeper on this issue let us next consider the Kerr metric.

\unskip
\subsection{Frame dragging in the Kerr metric}

The calculations performed in the previous subsection can be very easily repeated for the Kerr metric.

In  Boyer--Linquist coordinates  the Kerr metric takes the form
\begin{widetext}
\begin{eqnarray}
ds^2&=&\left(-1+\frac{2mr}{r^2+a^2\cos^2\theta}\right)dt^2-\left(\frac{4mar\sin^2\theta}{r^2+a^2\cos^\theta}\right)dtd\phi+\left(\frac{r^2+a^2\cos^2\theta}{r^2-2mr+a^2}\right)dr^2+(r^2+a^2\cos^2\theta)d\theta^2\nonumber \\&+&\left(r^2\sin^2\theta+a^2\sin^2\theta+\frac{2mra^2\sin^4\theta}{r^2+a^2\cos^2\theta}\right)d\phi^2,
\label{k1}
\end{eqnarray}
\end{widetext}
and the congruence of world--lines of observers at rest in (\ref{k1}) are defined by the time--like vector $u^\alpha$ with components

\begin{equation}
u^{\alpha}=\left(\frac{1}{\sqrt{1-\frac{2mr}{r^2+a^2\cos^2\theta}}},0,0,0\right).
\label{Ku}
\end{equation}
This congruence is endowed with vorticity, described by a vorticity vector $\omega^\alpha$ whose  non--vanishing components are
\begin{widetext}
\begin{eqnarray}
\omega^r=2mra\cos\theta(r^2-2mr+a^2)(r^2+a^2\cos^2\theta)^{-2}(r^2-2mr+a^2\cos^2\theta)^{-1},
\label{or}
\end{eqnarray}
\end{widetext}
and
\begin{widetext}
\begin{eqnarray}
\omega^\theta=ma\sin\theta(r^2-a^2\cos^2\theta)(r^2+a^2\cos^2\theta)^{-2}(r^2-2mr+a^2\cos^2\theta)^{-1}.
\label{othe}
\end{eqnarray}
\end{widetext}
The above expressions coincide with (\ref{w1}) and (\ref{w2}) up to first order in $m/r$ and $a/r$.

Finally,  using the package GR- Tensor running on Maple we obtain for the super--Poynting vector (\ref{supepo})
\begin{equation}
P^\mu=(P^t, 0, 0, P^\phi),
\label{pk}
\end{equation}
with
\begin{widetext}
\begin{eqnarray}
P^t&=&-18m^3ra^2\sin^2\theta(r^2-2mr+a^2\sin^2\theta+a^2)(r^2+a^2\cos^2\theta)^{-4}(r^2-2mr+a^2\cos^2\theta)^{-2}\nonumber \\&\times&\left(\frac{r^2-2mr+a^2\cos^2\theta}{r^2+a^2\cos^2\theta}\right)^{-1/2},
\label{pkt}
\end{eqnarray}
\begin{eqnarray}
P^\phi&=&9m^2a(r^2-2mr-a^2\cos^2\theta+2a^2)(r^2+a^2\cos^2\theta)^{-4}(r^2-2mr+a^2\cos^2\theta)^{-1}\nonumber \\&\times&\left(\frac{r^2-2mr+a^2\cos^2\theta}{r^2+a^2\cos^2\theta}\right)^{-1/2}.
\label{pkfi}
\end{eqnarray}
\end{widetext}
From the above expressions it follows, as in the precedent case, that there is  an azimuthal flow of superenergy as long as $a\neq0$, inversely the vanishing of such a flow implies $a=0$. Once again frame dragging appears to be tightly related to a circular flow of superenergy on the plane orthogonal to the vorticity vector.

In the present case we can delve deeper in the relationship between the source of the field and the vorticity, since a specific interior for the Kerr metric is available \cite{20c}. The remarkable fact 
is  the presence of a non--vanishing $T^\phi_t$ component of the energy--momentum tensor of the source, which, defining as usual an energy--momentum flux vector as: $F^\nu=-V^\mu T_{\nu \mu}$ (where $V^\mu$ denotes the four velocity of the fluid), implies   that  in the equatorial plane of our system  (within the source) energy flows round in circles around the symmetry axis. This result, as we shall see in the next section, is a  reminiscence of an effect appearing in stationary Einstein--Maxwell systems.  Indeed, in all stationary Einstein--Maxwell systems, there is a non vanishing component of the Poynting vector describing a similar phenomenon \cite{14, 15} (of electromagnetic nature, in this latter case). Thus, the appearance of such a component seems to be a distinct physical property of rotating fluids, which has been overlooked in previous studies of  these sources, and that is directly related to the vorticity (see eqs.(8) and (18) in \cite{20c}).

\subsection{Frame dragging in a general stationary vacuum spacetime}
Let us now consider the general stationary and axisymmetric vacuum case.

The line element for a general stationary and axisymmetric vacuum
spacetime may be written as \cite{21b, 22b}
\begin{equation}
ds^2=-fdt^2+2f\omega dt d\phi+f^{-1} e^{2
\gamma}(d\rho^2+dz^2)+(f^{-1}\rho^2-f\omega^2)d\phi^2,
\label{metrica}
\end{equation}
where $x^0=t$; $x^1=\rho$; $x^2=z$ and $x^3=\phi$ and metric
functions depend only on $\rho$ and $z$, which must satisfy the
vacuum field equations:
\begin{equation}\label{EFE1}
    \gamma_\rho = \frac{1}{4\rho f^2}\left[ \rho^2\left( f_\rho^2-f_z^2  \right)-f^4 \left( \omega_\rho^2-\omega_z^2  \right)   \right],
\end{equation}

\begin{equation}\label{EFE2}
    \gamma_z = \frac{1}{2\rho f^2}\left( \rho^2 f_\rho f_z - f^4 \omega_\rho\omega_z   \right),
\end{equation}

\begin{equation}\label{EFE3}
    f_{\rho \rho} = - f_{zz} - \frac{f_\rho}{\rho} -\frac{f^3}{\rho^2}\left( \omega_\rho^2+\omega_z^2 \right) + \frac{1}{f} \left(  f_\rho^2 +
    f_z^2\right),
\end{equation}

\begin{equation}\label{EFE4}
    \omega_{\rho \rho} = - \omega_{zz} + \frac{\omega_\rho}{\rho} -\frac{2}{f}\left(f_\rho \omega_\rho + f_z\omega_z \right),
\end{equation}
where subscripts denote partial derivatives.

Then following the same protocol as in the previous cases, we define the four velocity vector for an observer at rest in the frame of
(\ref{metrica}), which reads
\begin{equation}
u^{\alpha}=(f^{-1/2},0,0,0) \label{four}.
\end{equation}

The super-Poynting vector can now  be calculated for the general
class of spacetimes represented by the above metric (\ref{metrica}),
(i.e.: without making any assumption about the matter content of the
source), and one gets (using again GR--Tensor)
\begin{equation}\label{SP0}
    P^\mu = (P^t, 0, 0, P^\phi) \quad {\mathrm{with}} \quad P^t = \omega
    P^\phi, 
\end{equation}

where $P^\phi$ is given by (again in
the general case, i.e.: without taking into account the field
equations):
\begin{eqnarray}
  \nonumber P^\phi = f^{3/2}e^{-4\gamma}\rho^{-5} \left\{  A11  \right\},  
\label{P20}
\end{eqnarray}

or using the field equations  (\ref{EFE1}-\ref{EFE4})
in the above expression

\begin{eqnarray}
  \nonumber P^\phi = -\frac{1}{32}f^{-3/2}e^{-4\gamma}\rho^{-5} \left\{  A12  \right\},
\label{P20}
\end{eqnarray}
where $A11$ and $A12$ are given in the Appendix A.

Now,  in \cite{15} it has been shown that for the general
metric (\ref{metrica}) the following relations hold
\begin{equation} \label{ultima}
H_{\alpha \beta} = 0 \Leftrightarrow \omega^{\alpha} = 0
\Leftrightarrow \omega =0 .
\end{equation}
and of course as it follows from  (\ref{supepo})
\begin{equation} \label{ultima}
H_{\alpha \beta} = 0 \Rightarrow P^\mu = 0\, .
\end{equation}
In order to establish a link between vorticity and the super--Poynting vector of the kind already found for the Kerr (and Lense--Thirring) metric we still need to prove that the vanishing of the super--Poynting vector implies the vanishing of the vorticity, i.e. we have to prove that 
\begin{equation} \label{ultimab}
 P^\mu = 0 \Leftrightarrow H_{\alpha \beta} = 0 \Leftrightarrow \omega^{\alpha} = 0 \Leftrightarrow \omega =0\, .
\end{equation}

Such a  proof  has been carried out in \cite{13b}, but is quite cumbersome and therefore we shall omit the details here.

Thus based on (\ref{ultimab})  we conclude that for any stationary spacetime, irrespectively of its source, there is a frame dragging effect associated to a flux of superenergy on the plane orthogonal to the vorticity vector.

We shall next analyze the electro--vacuum stationary case.

\section{Frame dragging in electro--vacuum stationary spacetimes}

Electro--vacuum solutions to the Einstein equations pose a challenge concerning  the frame dragging effect.  This was pointed out for the first time by Bonnor in  \cite{14} by analyzing the gravitational field produced by a magnetic dipole with an electric charge in the center. The surprising result  is  that, for this spacetime, the world lines of observers at rest with respect to the electromagnetic source are endowed of vorticity (i.e. the resulting spacetime is not static but stationary).

In  order  to explain the appearance of vorticity in the spacetime generated by a charged magnetic dipole  Bonnor resorts to a result pointed out by Feynmann in his Lectures on Physics \cite{23bc}, showing  that for such a system (in the context of classical electrodynamics) there exists  a non--vanishing component of the Poynting vector describing a flow of electromagnetic energy round in circles. This strange  result leads Feynmann to write that  ``it shows the theory of the Poynting vector is obviously nuts''. However, some pages ahead in the same book, when discussing the ``paradox'' of the rotating disk with charges and a solenoid, Feynmann shows that this ``circular'' flow of electromagnetic energy is absolutely necessary  in order to preserve the conservation of angular momentum. In other words the theory of the Poynting vector  not only is not ``nuts'', but is necessary to reconcile the electrodynamics with the conservation law of angular momentum.

 Based on the above comments  Bonnor  then suggests that, in the context of GR,  such a circular flow of energy affects inertial frames by producing vorticity of congruences of particles, relative to the compass of inertia. In other words Bonnor suggests that the ``something'' which rotates thereby  generating the vorticity,  is electromagnetic energy.

 The interesting point is that this conjecture was shown to be valid for a general axially symmetric stationary electro--vacuum metric \cite{15}. 
 
 Indeed, assuming the line element (\ref{metrica}) for the spacetime admitting an electromagnetic field, it can be shown that  the variable responsible for the rotational multipole moments, which in its turn determine the vorticity of the congruence of world lines of observers,  is affected by, both, the electromagnetic field and by the mass rotations (angular momentum) \cite{15}. This explains why the vorticity does not necessarily vanish in the case when the angular momentum of the source is zero but electromagnetic fields are present. At any rate, it is important to stress that in such cases, the super--Poynting vector does not vanish either.
 
 We shall next consider the presence of vorticity  due to gravitational and electromagnetic radiation.
  \section{Vorticity and radiation}
We shall now analyze the generation of vorticity related to the emission of  gravitational  and/or electromagnetic radiation. As we shall see, the emission of radiation  is always accompanied by the appearance of vorticity of world lines of observers. Furthermore, the calculations  suggest that  once  the radiation process has stopped, there is still a remaining vorticity associated to  the tail of the wave, which allows in principle to prove (or disprove) the violation of the Huyghens principle in a Riemannian spacetime (see \cite{41, 42, 43, 44, 45, 46, 47} and references therein for a discussion on this issue),  by means of observations.
\subsection{Gravitational radiation and vorticity}

Since the early days of GR, starting with the works of Einstein and Weyl on the linear approximation  of the Einstein equations, a great deal of work has been done so far in order to provide
a consistent framework for the study of gravitational radiation. Also, important collaboration efforts have been carried on, and are now under consideration, to put in evidence gravitational waves by means of  laser  interferometers \cite{48}.

However it was necessary to wait for more than half a century, until Bondi and coworkers \cite{14n}  provided a firm theoretical evidence of the existence of gravitational radiation without resorting to the linear approximation.

The essential ``philosophy''  behind the Bondi formalism, consists in interpreting gravitational radiation as  the physical process by means of which the source of the field ``informs'' about any changes in its structure. Thus the information required to forecast the evolution of the system (besides the ``initial'' data) is thereby identified  with radiation itself, and   this information is represented by the so called ``news function''. In other words, whatever happens at the source, leading to changes in the field, it can only
do so by affecting the news function and vice versa. Therefore if the news function is zero over a time interval, there is no gravitational radiation over that interval. Inversely, non vanishing news on an interval implies the emission  of gravitational radiation during that interval. Thus the  main virtue  of this approach resides in  providing a clear and precise
criterion for the existence of gravitational radiation.

The above described  picture is reinforced by the fact that the Bondi mass of a system is constant if and only if there are no news.

In order to facilitate  discussion  let us briefly introduce the main aspects of the Bondi approach.
Bondi and coworkers start with the general form of an axially (and reflection) symmetric asymptotically flat
metric given by
\begin{eqnarray}
ds^2 & = & \left(\frac{V}{r} e^{2\beta} - U^2 r^2 e^{2\gamma}\right) du^2
+ 2 e^{2\beta} du dr \nonumber \\
& + & 2 U r^2 e^{2\gamma} du d\theta
- r^2 \left(e^{2 \gamma} d\theta^2 + e^{-2\gamma} \sin^2{\theta} d\phi^2\right),
\label{Bm}
\end{eqnarray}
where $V, \beta, U$ and $\gamma$ are functions of
$u, r$ and $\theta$.

The coordinates are numbered $x^{0,1,2,3} = u, r, \theta, \phi$ respectively.
$u$ is a timelike coordinate such that $u=constant$ defines a null surface.
In flat spacetime this surface coincides with the null light cone
open to the future. $r$ is a null coordinate ($g_{rr}=0$) and $\theta$ and
$\phi$ are two angle coordinates.

Regularity conditions in the neighborhood of the polar axis
($\sin{\theta}=0$), implies that
as $\sin{\theta}->0$
\begin{equation}
V, \beta, U/\sin{\theta}, \gamma/\sin^2{\theta},
\label{regularity}
\end{equation}
each equals a function of $\cos{\theta}$ regular on the polar axis.

Then the  four metric functions are assumed to be expanded in series of $1/r$,
which  using field equations produces

\begin{equation}
\gamma = c(u,\theta) r^{-1} + \left[C(u,\theta) - \frac{1}{6} c^3\right] r^{-3}
+ ...,
\label{ga}
\end{equation}
\begin{equation}
U = - \left(c_\theta + 2 c \cot{\theta}\right) r^{-2} + ...,
\label{U}
\end{equation}
\begin{widetext}
\begin{eqnarray}
V  =  r - 2 M(u,\theta)
 -  \left[ N_\theta + N \cot{\theta} -
c_{\theta}^{2} - 4 c c_{\theta} \cot{\theta} -
\frac{1}{2} c^2 (1 + 8 \cot^2{\theta})\right] r^{-1} + ...,
\label{V}
\end{eqnarray}
\end{widetext}
\begin{equation}
\beta = - \frac{1}{4} c^2 r^{-2} + ...
\label{be}
\end{equation}
where letters as subscripts denote derivatives, and

\begin{equation}
4C_u = 2 c^2 c_u + 2 c M + N \cot{\theta} - N_\theta.
\label{C}
\end{equation}

The three functions $c, M$ and $N$ depend on $u$ and $\theta$, and are further
related by the supplementary conditions
\begin{equation}
M_u = - c_u^2 + \frac{1}{2}
\left(c_{\theta\theta} + 3 c_{\theta} \cot{\theta} - 2 c\right)_u,
\label{M}
\end{equation}
\begin{equation}
- 3 N_u = M_\theta + 3 c c_{u\theta} + 4 c c_u \cot{\theta} + c_u c_\theta.
\label{N}
\end{equation}

In the static case $M$ equals the mass of the system whereas $N$ and $C$
are closely related to the dipole and quadrupole moment respectively.

Next, Bondi defines the mass $m(u)$ of the system as
\begin{equation}
m(u) = \frac{1}{2} \int_0^\pi{M(u,\theta) \sin{\theta} d\theta},
\label{m}
\end{equation}
which by virtue of (\ref{M}) and (\ref{regularity}) yields
\begin{equation}
m_u = - \frac{1}{2} \int_0^\pi{c_u^2 \sin{\theta} d\theta}.
\label{mub}
\end{equation}

The two main conclusions emerging from the Bondi's approach are
\begin{itemize}
\item If $\gamma, M$ and $N$ are known for some $u=a$(constant), and
$c_u$ (the news function) is known for all $u$ in the interval
$a \leq u \leq b$,
then the system is fully determined in that interval. 
\item As it follows from (\ref{mub}), the mass of a system is constant
if and only if there are no news.
\end{itemize}
In the light of these comments the relationship between news function
and the occurrence of radiation becomes clear.

Let us now  calculate the vorticity of the world lines of observers at rest in the frame of (\ref{Bm}).
For such observers the four-velocity
vector has components
\begin{equation}
u_\alpha = \left(A, \frac{e^{2\beta}}{A}, \frac{U r^2 e^{2\gamma}}{A}, 0\right)
\label{fv}
\end{equation}
with
\begin{equation}
A \equiv \left(\frac{V}{r} e^{2\beta} - U^2 r^2 e^{2\gamma}\right)^{1/2}.
\label{A}
\end{equation}
Using (\ref{vv}) , we easily obtain
\begin{equation}
\omega^\alpha = \left(0, 0, 0, \omega^\phi \right)
\label{oma}
\end{equation}
with
\begin{widetext}
\begin{eqnarray}
\omega^\phi  = -\frac{e^{-2\beta}}{2 r^2 \sin{\theta}}
\{
2 \beta_\theta e^{2\beta}
- \frac{2 e^{2\beta} A_\theta}{A}
- \left(U r^2 e^{2\gamma}\right)_r 
 +  \frac{2 U r^2 e^{2\gamma}}{A} A_r
+ \frac{e^{2\beta} \left(U r^2 e^{2\gamma}\right)_u}{A^2}
- \frac{U r^2 e^{2\gamma}}{A^2} 2 \beta_u e^{2\beta}
\}
\label{om3}
\end{eqnarray}
\end{widetext}
and for the absolute value of $\omega^\alpha$ we get
\begin{eqnarray}
\Omega & \equiv & \left(- \omega_\alpha \omega^\alpha\right)^{1/2} =
 \frac{e^{-2\beta -\gamma}}{2 r}
\{2 \beta_\theta e^{2\beta} - 2 e^{2\beta} \frac{A_\theta}{A}
 -  \left(U r^2 e^{2\gamma}\right)_r
\nonumber \\
& + & 2 U r^2 e^{2\gamma} \frac{A_r}{A}
 +  \frac{e^{2\beta}}{A^2} \left(U r^2 e^{2\gamma}\right)_u
 - 2 \beta_u \frac{e^{2\beta}}{A^2} U r^2 e^{2\gamma}
\}
\label{OM}
\end{eqnarray}

Feeding back (\ref{ga})--(\ref{be}) into (\ref{OM}) and
keeping only  terms up to order $\frac {1}{r^2}$, we obtain
\begin{widetext}
\begin{eqnarray}
\Omega  = -\frac{1}{2r} ( c_{u \theta}+2 c_u \cot \theta) 
 +\frac 1{r^2} \left[ M_{\theta}-M (c_{u \theta}+2 c_u \cot \theta)-c c_{u
\theta}+6 c c_u \cot \theta+2 c_u c_{\theta} \right].
\label{Om2}
\end{eqnarray}
\end{widetext}
Let us now analyze the expression above. First of all observe that, up to order $1/r$, a gyroscope in the gravitational field given by (\ref{Bm}) will precess as long as
the system radiates ($c_{u} \not= 0$).
Indeed, if we assume
\begin{equation}
c_{u\theta} + 2 c_u \cot{\theta} = 0
\label{if}
\end{equation}
then
\begin{equation}
c_u = \frac{F(u)}{\sin^2{\theta}}
\label{cu}
\end{equation}
which implies, due to the regularity conditions (\ref{regularity})
\begin{equation}
F(u) = 0  \Longrightarrow c_u = 0.
\label{F}
\end{equation}
In other words the leading term in (\ref{Om2}) will vanish if and only if $c_u = 0$.

Let us now analyze the term of order ${\displaystyle \frac1{r^2}}$. It   contains, besides the
terms involving $c_u$, a term not involving news
($M_{\theta}$). Let us now assume that initially (before some $u = u_0 =
$constant) the system is static, in which case
\begin{equation}
c_u = 0
\label{static}
\end{equation}
which implies , because of (\ref{N})
\begin{equation}
M_{\theta} = 0
\label{staticII}
\end{equation}

and $\Omega = 0$ (actually, in this case $\Omega=0$ at any order) as expected
for  a static field.  Then let us suppose that at $u = u_0$ the system starts to
radiate ($c_u \neq 0$) until $u = u_f$, when the news function vanishes again. For $u>u_f$
the system is not radiating although (in general) $M_{\theta} \neq 0$ implying
time dependence of metric functions. This class of spacetimes is referred to as non-radiative motions \cite{14n}.

Thus, in the interval $u \in$ ($u_0$,$u_f$) the leading term of vorticity is given by the term of the order $1/r$ in  (\ref{Om2}).
For $u>u_f$ there is a vorticity term of order $\frac{1}{r^2}$
describing the effect of the tail of the wave on the vorticity. This 
provides  an ``observational'' possibility to find evidence for the violation of the Huygens's
principle.

Following the line of arguments of the preceding sections, we shall establish a link between vorticity and a circular flow of superenergy on the plane orthogonal to the vorticity vector.  For doing so, let us calculate   the  super--Poynting vector ($P^{\mu}$), defined by (\ref{supepo}). We obtain that the leading terms for
each super--Poynting component
are
\begin{equation}
P_r=-\frac{2}{r^2}c^2_{uu},\label{point1}
\end{equation} 
\begin{widetext}
\begin{eqnarray}
P_\theta =-\frac{2}{r^2 \sin\theta}\{
[2c_{uu}^2 c
+c_{uu}c_{u}]\cos\theta 
+\left[c_{uu}c_{\theta
u}+c_{uu}^2 c_{\theta}\right]\sin\theta \},\label{p2}
\end{eqnarray}
\end{widetext}
\begin{eqnarray}
P_\phi=P^{\phi}=0.
 \label{p3}
\end{eqnarray}

The vanishing (at all orders) of the  azimuthal component ($P^{\phi}$), is expected from the reflection symmetry
of the Bondi metric, which is incompatible with the presence of a circular flow of superenergy  in the $\phi$ direction. Since the vorticity vector, which is orthogonal to the plane of rotation,
has in the Bondi spacetime only one non--vanishing contravariant component ( $\phi$ ), then  the plane of the associated rotation is  orthogonal to the
$\phi$ direction. Therefore, it is  the $\theta$ component of $P^{\mu}$  the physical factor to be associated  to the  vorticity, in the Bondi case.

In order to strength  further the case for the super-Poynting vector as the physical origin of the mentioned vorticity, we shall consider next  the general radiative metric
without axial and reflection symmetry.
 
The extension of the Bondi formalism to the case without any kind of symmetries  was performed by Sachs \cite{43b}. In this  case the line element reads (we have found
more convenient to follow the notation given in \cite{Burg} which
is slightly different from the
original Sachs
paper)
\begin{widetext}
\begin{eqnarray}
ds^2&=&(Vr^{-1}e^{2\beta}-r^2 e^{2\gamma}U^2\cosh
2\delta -r^2 e^{-2\gamma} W^2
 \cosh 2\delta -2r^2 UW\sinh 2\delta)du^2
 +2e^{2\beta}dudr +2r^2(e^{2\gamma}U\cosh 2\delta +W \sinh
2\delta)dud\theta \nonumber \\
 &+&2r^2(e^{-2\gamma}W\cosh 2\delta 
 +U\sinh2\delta)\sin\theta dud\phi-r^2(e^{2\gamma}\cosh 2\delta d\theta^2
+e^{-2\gamma}\cosh 2\delta \sin^2\theta d\phi^2 
 +2\sinh2\delta \sin\theta d\theta d\phi),\label{bsm}
\end{eqnarray}
\end{widetext}
where $\beta$,
$\gamma$, $\delta$, $U$, $W$, $V$ are functions of $x^0=u$,
$x^1=r$,
$x^2=\theta$, $x^3=\phi$. 

The general analysis of the field
equations is similar to  the one  in
\cite{14n}, but of course the expressions
are far more complicated (see
\cite{43b, Burg} for details).
In
particular, there are now two news functions.

Let us
first calculate the vorticity for the congruence
of observers at rest in
(\ref{bsm}), whose four--velocity vector is given
by  
\begin{equation}
u^{\alpha} =A^{-1}\delta^{\alpha}_{u},
\label{ug}
\end{equation}
where now $A$ is given
by
\begin{widetext}
\begin{eqnarray}
A=(Vr^{-1}e^{2\beta}-r^2 e^{2\gamma}U^2\cosh 2\delta
-r^2 e^{-2\gamma} W^2
\cosh 2\delta -2r^2 UW\sinh
2\delta)^{1/2}.
\end{eqnarray}
\end{widetext}
Thus, (\ref{vv}) lead us
to
\begin{equation}
\omega^\alpha=(\omega^u,\omega^r,\omega^\theta,\omega^\phi),
\end{equation}

where
\begin{widetext}
\begin{eqnarray}\label{vor1}
\omega^u&=&-\frac{1}{2A^2\sin\theta}
\{r^2e^{-2\beta}(WU_{r}-UW_{r})+
\left[2r^2\sinh 2\delta
\cosh 2\delta (U^2e^{2\gamma}+W^2e^{-2\gamma})+
4UWr^2\cosh^2
2\delta\right]e^{-2\beta}\gamma_{r}\nonumber
\\
&+&2r^2e^{-2\beta}(W^2e^{-2\gamma}
-U^2e^{2\gamma})\delta_{r}
+e^{2\beta}[e^{-2\beta}(U\sinh
2\delta+e^{-2\gamma} W\cosh
2\delta)]_{\theta}\nonumber \\&-&
e^{2\beta}[e^{-2\beta}(W\sinh
2\delta+e^{-2\gamma} U\cosh
2\delta]_{\phi}\},
\end{eqnarray}
\end{widetext}
\begin{widetext}
\begin{eqnarray}\label{vor2}
\omega^r&=&\frac{1}{e^{2\beta}\sin\theta}\{2r^2 A^{-2}[ (
(U^2e^{2\gamma}+W^2e^{-2\gamma})\sinh
2\delta \cosh 2\delta+
UW\cosh^2 2\delta) \gamma_{u}
+(W^2e^{-2\gamma}-U^2e^{2\gamma})\delta_{u}
+\frac{1}{2}(WU_{u}-UW_{u})]\nonumber \\
&+&A^2[A^{-2}(We^{-2\gamma}\cosh 2\delta+U\sinh 2\delta)]_{\theta}
-
A^2[A^{-2}(W\sinh 2\delta+Ue^{2\gamma}\cosh
2\delta)]_{\phi}\},
\end{eqnarray}
\end{widetext}
\begin{widetext}
\begin{eqnarray}\label{vor3}
\omega^\theta&=&\frac{1}{2
r^2\sin\theta}\{A^2e^{-2\beta}[r^2A^{-2}(U\sinh
2\delta+We^{-2\gamma}\cosh
2\delta)]_{r}\nonumber
\\ &-&e^{2\beta} A^{-2}[e^{-2\beta} r^2(U\sinh
2\delta+e^{-2\gamma} W\cosh
2\delta)]_{u}+e^{2\beta} A^{-2}(e^{-2\beta}
A^2)_{\phi}
\},
\end{eqnarray}
\end{widetext}
and
\begin{widetext}
\begin{eqnarray}\label{vor4}
\omega^\phi&=&\frac{1}{2r^2\sin\theta}
\{
A^2e^{-2\beta}[r^2 A^{-2} (W\sinh 2\delta+Ue^{2\gamma}\cosh
2\delta)]_{r}
\nonumber \\
&-&e^{2\beta} A^{-2}[r^2e^{-2\beta}(W\sinh
2\delta+Ue^{2\gamma}\cosh
2\delta)]_{u}+A^{-2}e^{2\beta}
(A^2e^{-2\beta})_{\theta}\}.
\end{eqnarray}
\end{widetext}
Expanding the metric functions  in series of $1/r$ as in the previous case, using the field equations and feeding back the resulting expressions into (\ref{vor1}, \ref{vor2}, \ref{vor3}, \ref{vor4})
  we get for the leading term of
the absolute value of $\omega^\mu$
\begin{widetext}
\begin{eqnarray}
\Omega=-\frac{1}{2r}[(c_{\theta u}+2c_{u}\cot\theta +
d_{\phi u} \csc\theta)^2
+(d_{\theta u}+2d_{u}\cot\theta - c_{\phi u}
\csc\theta)^2]^{1/2},\label{On}
\end{eqnarray}
\end{widetext}
which of course reduces to
(\ref{Om2}) in the Bondi (axially and reflection
symmetric) case
($d=c_{\phi}=0$).
It is worth stressing the fact that now we have two news functions ($c_u, d_u$).

Next, the calculation of the super--Poynting  vector gives the
following
result
\begin{equation}
P_\mu=(0,P_r,P_\theta,P_\phi).
\end{equation}
The explicit terms are too long and the calculations are quite cumbersome (see \cite{19} for details), so let us just present the leading terms for
each super--Poynting component, they read

\begin{equation}
P_r=-\frac{2}{r^2}(d^2_{uu}+c^2_{uu}),\label{point1}
\end{equation} 
\begin{widetext}
\begin{eqnarray}
P_\theta =-\frac{2}{r^2 \sin\theta}\{
[2(d_{uu}^2+c_{uu}^2)c
+c_{uu}c_{u}+d_{uu}d_{u}]\cos\theta +\left[c_{uu}c_{\theta
u}+d_{uu}d_{\theta
u}+(c_{uu}^2+d_{uu}^2)c_{\theta}\right]\sin\theta
+\nonumber \\
+c_{uu}d_{\phi u}-d_{uu}c_{\phi u}
+(d_{uu}^2+c_{uu}^2)d_{\phi}\},\label{p2}
\end{eqnarray}
\end{widetext}
\begin{widetext}
\begin{eqnarray}
P_\phi=\frac
{2}{r^2}\{ 2[c_{uu}^2d_{u}-d_{uu}c_{u}-(d_{uu}^2+c_{uu}^2)d]\cos\theta +
\nonumber \\
+\left[c_{uu}d_{\theta
u}-d_{uu}c_{\theta
u}-(c_{uu}^2+d_{uu}^2)d_{\theta}\right]\sin\theta
+(d_{uu}^2+c_{uu}^2)c_{\phi}-(c_{uu}c_{\phi u}+d_{uu}d_{\phi
u}) \} \label{p3}
\end{eqnarray}
\end{widetext}
from which it follows
\begin{widetext}
\begin{eqnarray}
P^{\phi} =
-\frac{2}{r^4 \sin^2\theta}\{ \sin\theta \left[ d_{u\theta}
c_{uu} - d_{uu}
c_{u \theta} \right] + 2\cos\theta \left[c_{uu} d_{u} -
d_{uu} c_{u}
\right]- [c_{uu}
c_{u\phi}+ d_{uu}d_{u \phi}]\}
,\label{fi}
\end{eqnarray}
\end{widetext}
this component of course vanishes in the Bondi case.

From the expressions above we see that the main conclusion established for the Bondi metric, is also  valid in the most general case,  namely,
  there is always a non--vanishing component of $P^\mu$ on the plane orthogonal to a unit vector along which there is a non--vanishing component of
vorticity, and  inversely,   $P^\mu$ vanishes on a plane orthogonal to a unit vector along which the component of
vorticity vector vanishes. The link between the super--Poynting vector and vorticity is thereby firmly established.

So far we have shown  the appearance of vorticity in stationary vacuum spacetimes,  stationary electro--vacuum spacetimes and  in radiative vacuum spacetimes (Bondi--Sachs), and have succeed in exhibiting  the link between this vorticity  and a circular flow of  electromagnetic and/or super--energy,  on the plane orthogonal to the vorticity vector. It remains to analyze the possible role of electromagnetic radiation in the appearance  of vorticity. The next section is devoted to this issue.

\subsection{Electromagnetic radiation and vorticity}
The relationship between electromagnetic radiation and vorticity has been unambiguously established in \cite{49}. The corresponding calculations are quite cumbersome and we shall not reproduce them here. Instead  we shall  highlight the most important results emerging from such calculations.

The formalism used to study the general electro--vacuum case (including electromagnetic radiation) was developed by  van der Burg in \cite{50}.  It represents a generalization of the Bondi--Sachs formalism for the  Einstein--Maxwell system.

Thus, the  starting point is the Einstein--Maxwell system of equations,  which reads
\begin{eqnarray}
R_{\mu\gamma} + T_{\mu\gamma} = 0,\\
F_{[\mu\nu,\delta]}=0,\\
F^{\mu\nu}_{;\nu}=0,
\end{eqnarray}
where $R_{\mu\gamma}$ is the Ricci tensor and the energy momentum tensor $T_{\mu\gamma}$ of the electromagnetic field is given as usual by
\begin{equation}
T_{\mu\nu}=\frac{1}{4}g_{\mu\nu}F_{\gamma\delta}F^{\gamma\delta}-g^{\gamma\delta}F_{\mu\gamma}F_{\nu\delta}.
\end{equation}

Then, following the script indicated in \cite{14n}, i.e. expanding the physical and metric variables in power series  of $1/r$ and using the Einstein--Maxwell equations, one arrives at the conclusion that if a specific set of functions is  prescribed on a given initial hypersurface $u=constant$, the evolution of the system is fully determined provided the four functions,  $c_u, d_u, X, Y$ are given for all $u$. These four functions are the news functions of the system. The first two ($c_u, d_u$ ) are the gravitational news functions already mentioned before  for the purely gravitational case, whereas $X$ and $Y$ are the two news functions corresponding to the  electromagnetic field, these appear in the series expansion of $F_{12}, F_{13}$. Thus, whatever happens at the source leading to changes in the field, it can only do so by affecting the four news functions and viceversa. 

Following the same line of arguments, an equation for the decreasing of the mass function  due to the radiation (gravitational and electromagnetic) similar to (\ref{mub}) can be obtained, it reads
\begin{equation}
m_u=-\int^{2\pi}_0\int^\pi_0( c^{\ast}_u \bar c^{\ast}_u+\frac{1}{2}X^{\ast} \bar X^{\ast})\sin\theta d\theta d\phi,
\label{mas5}
\end{equation}
where 
\begin{equation}
c^{\ast}=c+id, \qquad X^{\ast}=X+i Y,
\label{m4}
\end{equation}
and bar denotes complex conjugate.

Having arrived at this point we can now proceed to calculate the vorticity, the super--Poynting vector and the electromagnetic Poynting vector. The resulting expressions are available in \cite{49}, since they are extremely long, here we shall focus on the main conclusions emerging from them.

First, the vorticity vector  (\ref{vv}) is calculated for the four--vector $u^\alpha$ given by (\ref{ug}). The important point to stress here is that the absolute value of $\omega^\mu$ can be  written  generically as 
\begin{eqnarray}\label{ome}
\Omega&=&\Omega_\mathcal{G}r^{-1}+\cdots+\Omega_{\mathcal{GEM}}r^{-3}+\cdots,
\end{eqnarray}
where subscripts $\mathcal{G}$, $\mathcal{GEM}$ and $\mathcal{EM}$ stand for gravitational, gravito--electromagnetic  and electromagnetic. The   ``gravitational''  subscript  refers to those terms containing exclusively functions appearing in the purely gravitational case and their derivatives. ``Electromagnetic'' terms are those containing exclusively  functions  appearing in $F_{\mu \nu}$  and their derivatives, whereas ``gravito--electromagnetic'' subscript  refers to those terms containing functions of either kind and/or combination of both.

Finally, we calculate the electromagnetic Poynting vector defined by
\begin{equation}
S^\alpha=T^{\alpha\beta}u_\beta,
\end{equation}
and the super--Poynting vector defined by (\ref{supepo}). Since we are not operating in vacuum, $P(C)_\alpha$ and $P(R)_\alpha$ are different, we shall use  $P(C)_\alpha$ for the discussion.

The resulting expressions are deployed in \cite{49}. Let us summarize the main information contained in such expressions.

First, we notice that the leading terms for each super--Poynting (contravariant) component are
\begin{eqnarray}
P^u&=&P^{u}_{\mathcal{G}}r^{-4}+\cdots, \nonumber\\ \label{sp0}
P^r&=&P^r_{\mathcal{G}}r^{-4}+\cdots, \nonumber\\ \label{sp1}
P^\theta &=& P^{\theta}_{\mathcal{G}}r^{-4} +\cdots+ P^{\theta}_{\mathcal{GEM}}r^{-6}+\cdots, \nonumber\\\label{sp2}
P^\phi &=& P^{\phi}_{\mathcal{G}}r^{-4} +\cdots+ P^{\phi}_{\mathcal{GEM}}r^{-6}+\cdots, \label{sp3}
\end{eqnarray}
whereas for  the electromagnetic Poynting vector we can write
\begin{equation}
S^u=S^u_\mathcal{EM}r^{-4}+S^u_\mathcal{GEM}r^{-5}\cdots,
\end{equation}
\begin{equation}
S^r=S^r_\mathcal{EM}r^{-2} + S^r_\mathcal{GEM} r^{-3}+\cdots,
\end{equation}
\begin{equation}
S^\theta=S^\theta_\mathcal{EM} r^{-4}+S^\theta_\mathcal{GEM} r^{-5}+\cdots,
\end{equation}
\begin{equation}
S^\phi=S^\phi_\mathcal{EM}r^{-4}+S^\phi_\mathcal{GEM} r^{-5}+\cdots.
\end{equation}

Next, there are explicit contributions from the electromagnetic news functions in  $\Omega_{\mathcal{GEM}}$ as well as in $P^{\phi}_{\mathcal{GEM}}$ and  $P^{\theta}_{\mathcal{GEM}}$. More so, the vanishing of  these contributions in  $P^{\phi}_{\mathcal{GEM}}$ and $ P^{\theta}_{\mathcal{GEM}}$ implies the vanishing of the corresponding contribution in $\Omega_{\mathcal{GEM}}$, and viceversa.

From the above it is clear that   electromagnetic radiation as described by electromagnetic news functions does produce vorticity.  Furthermore we have identified the presence of electromagnetic news both in the Poynting and the super--Poynting components orthogonal to the vorticity vector. Doing so we have proved that a Bonnor--like mechanism to generate vorticity is at work in this case too, but with the important difference that now vorticity  is generated by the contributions of, both,  the Poynting   and  the super--Poynting vectors,   on the planes orthogonal to the vorticity vector.

\section{Discussion}
We started this manuscript with two goals in mind. On the one hand we wanted to identify the fundamental physical phenomenon which being present in all scenarios exhibiting frame dragging, could be considered as the responsible for the frame dragging effect. In other words we wanted to identify the factor that mediates between the source of the gravitational field and the appearance of vorticity, in any scenario.

On the other hand, we wanted to explore the  observational consequences  that could be derived from our analysis.

Concerning our first goal, it has been clearly established that in vacuum, the appearance of vorticity is always related to the existence of  circular flow of super--energy in the plane orthogonal to the vorticity vector. This is true for all stationary vacuum  spacetimes as well as for general Bondi--Sachs  radiative spacetimes.

In the case of electro--vacuum  spacetimes, we have circular flows of super--energy as well as circular flows of electromagnetic  energy in the plane orthogonal to the vorticity vector. This is true in stationary electro--vacuum spacetimes as well as in spacetimes admitting, both, gravitational and electromagnetic radiation. Particularly remarkable is the fact that electromagnetic radiation does produce vorticity.

All this having been said, a natural question arises concerning our second goal, namely, what observational consequences could be derived from the analysis presented so far?

First of all it should be clear  that the established fact that the emission of  gravitational radiation always entails the appearance of vorticity in the congruence of the world lines of observers, provides a mechanism for detecting gravitational radiation.  Thus, any experimental device intended to measure  rotations could be  a potential detector of gravitational radiation as well. We  are well aware of the fact that extremely high sensitivities have to be reached, for these detectors to be operational. Thus,  from the estimates displayed in \cite{16}, we see that for a large class of possible events leading to the emission of gravitational radiation, the expected values of $\Omega$ range from $\Omega\approx  10^{-15}s^{-1}$  to $\Omega\approx 10^{-19}s^{-1}$. Although these estimates are twenty years old and deserves to be updated, we  believe that  probably the  sensitivity of  the actual technology  is still below the range of expected values of vorticity. 
Nevertheless, the  intense activity deployed in recent years in this field,  invoking  ring lasers,  atom interferometers, atom lasers,  anomalous spin--precession, trapped atoms  and  quantum interference (see References \cite{1g, 2g, 3g, 4g, 5g, 6g, 7g, 8g, 9g, 10g, 11g, 12g, 13g} and references therein), besides the incredible sensitivities obtained so far  in  gyroscope technology and exhibited in the Gravity Probe B experiment \cite{1}, make us being optimist  in that this kind of detectors may be operating in the foreseeable future.

In the same order of ideas  the established  link between vorticity and electromagnetic radiation, has  potential observational consequences which should not be overlooked.   Indeed, intense electromagnetic outbursts are expected  from hyperenergetic phenomena such as collapsing hypermassive neutron stars and Gamma Ray Bursts (see \cite{21x} and references therein).  Therefore, although the contributions of the ${\mathcal{GEM}}$ terms in (\ref{ome}) are of order $1/r^3$, in contrast with  the $\mathcal{G}$ terms  which are of order $1/r$, the coefficient of the former terms usually exceeds the latter by many orders of magnitude, which opens the possibility to detect them more easily.  

Finally, the association of the sources of electromagnetic fields (charges and currents) with vorticity,  suggests the possibility to extract information about the former, by measuring the latter. Thus, in \cite{14}, using the data corresponding  to  the earth, Bonnor estimates that the vorticity would be of the order of  $\Omega\approx 4\times 10^{-33}s^{-1}$. Although this figure is so small that we do not expect to be able to measure it in the near future, the strength of electromagnetic sources in very compact objects  could produce vorticity many orders of magnitude larger. 

To summarize. If we adopt the usual meaning of the verb ``to explain'' (a phenomenon), as referred to the action of  expressing   such a phenomenon in terms of fundamental concepts, then we can say that we have succeed in explaining the frame dragging effect as  due to circular flows of super--energy and electromagnetic energy (whenever present) in planes orthogonal to vorticity vector. This result in turn, creates huge opportunities to obtain information from self--gravitating systems by measuring the vorticity of the congruence of world--lines of  observers.
\vspace{6pt}

\begin{acknowledgments}
This research was funded by  Ministerio de Ciencia, Innovacion y Universidades. Grant number: PGC2018-- 096038--B--I00.
\end{acknowledgments}

\appendix
\section{}
\begin{eqnarray}
A11 &=& \left[-2\rho \omega_\rho(\omega_\rho^2 + \omega_z^2)\gamma_\rho -2 \rho\omega_z (\omega_\rho^2 + \omega_z^2)\gamma_z -\omega_z^2\omega_{\rho\rho}\rho
\right. 
\nonumber\\
&&\left.+ \omega_z^2\omega_{zz}\rho + 4\omega_z\omega_\rho\omega_{\rho z}\rho - \omega_\rho^2 \omega_{zz}\rho - \omega_\rho^3 + \omega_\rho^2\omega_{\rho\rho}\rho\right]f^4 
\nonumber\\
&&+3\rho(\omega_\rho^2 + \omega_z^2)(\omega_zf_z + \omega_\rho f_\rho)f^3 -2\rho (-2\rho\gamma_z\omega_{\rho z}+ 2\gamma_z^2 \omega_\rho \rho 
\nonumber\\
&&+2\rho\gamma_\rho^2\omega_\rho + \gamma_z\omega_z + \gamma_\rho\omega_\rho + \rho\gamma_\rho\omega_{zz} - \rho \gamma_\rho\omega_{\rho\rho}) f^2
\nonumber\\
&&\left[4\rho^3(f_z\omega_z + f_\rho\omega_\rho)\gamma_\rho^2 + 2\rho^2 (\rho f_{zz} \omega_\rho - 2\rho f_{\rho z} \omega_z - 2f_z\omega_z \right.
\nonumber\\
&&\left. + 4f_\rho \omega_\rho - 2\rho f_z \omega_{\rho z} - \rho f_{\rho\rho}\omega_\rho - \rho f_\rho \omega_{\rho\rho} + \rho f_\rho \omega_{zz})\gamma_\rho \right.
\nonumber\\
&&\left. + 4\rho^3(f_z\omega_z + f_\rho \omega_\rho)\gamma_z^2 + 2\rho^2(4f_\rho \omega_z + \rho f_z \omega_{\rho\rho} - 2\rho f_{\rho z} \omega_\rho \right. 
\nonumber\\
&& \left. + \rho f_{\rho\rho}\omega_z - 2\rho f_\rho \omega_{\rho z} - \rho f_{zz} \omega_z - \rho f_z \omega_{zz} + 2\omega_\rho f_z)\gamma_z\right.
\nonumber\\
&& \left. +4\rho^3 f_{\rho z} \omega_{\rho z} - \rho^3 f_{zz} \omega_{\rho\rho} - \rho^3 f_{\rho\rho}\omega_{zz} + \rho^2 f_{zz} \omega_\rho - 2 \rho^2 f_{\rho z} \omega_z\right.
\nonumber\\
&& \left. -\rho^2 f_{\rho\rho} \omega_\rho + \rho^3 f_{zz} \omega_{zz} + \rho^3 f_{\rho\rho} \omega_{\rho\rho}\right] f - 6\rho^3(f_\rho^2 + f_z^2)\omega_\rho\gamma_\rho
\nonumber \\
&& -6\rho^3(f_\rho^2 + f_z^2)\omega_z \gamma_z + 3 \rho^3 (f_{\rho\rho} f_\rho \omega_\rho + f_{zz} f_z \omega_z + 2f_{\rho z} f_z \omega_\rho 
\nonumber\\
&&- f_{\rho\rho} f_z \omega_z 
+2 f_{\rho z} f_\rho \omega_z - f_{zz} f_\rho \omega_\rho),
\nonumber
\end{eqnarray}

\begin{eqnarray}
A12 &=& \omega_\rho (-7 \omega_z^4- 6 \omega_\rho^2 \omega_z^2 + \omega_\rho^4) f^9 
\nonumber\\
&& + \left[-\rho \omega_\rho f_\rho ( \omega_z^4 + \omega_\rho^4 + 2 \omega_\rho^2 \omega_z^2 ) - \rho f_z \omega_z (\omega_\rho^4 + 2 \omega_z^2 \omega_\rho^2 +\omega_z^4)
\right]f^8
\nonumber \\
&& + \left[-4 \rho \omega_{zz}(\omega_\rho^2 + 3\omega_z^2) + 4 \omega_\rho (-2 \rho \omega_z \omega_{\rho z} + \omega_z^2)\right] f^7 
\nonumber\\
&& + \left[ 4 \rho \omega_z(-8 \omega_z^2 f_z + \rho \omega_z^2 f_{\rho z} -3 \rho \omega_\rho \omega_z f_{zz} -3 \rho \omega_\rho^2 f_{\rho z}-\rho f_\rho \omega_z \omega_{zz} \right.
\nonumber\\ 
&& \left.  - 5 \omega_\rho^2 f_z - 2 \rho f_\rho \omega_\rho \omega_{\rho z} -2 \omega_z \omega_\rho f_\rho 
- 2 \rho f_z \omega_\rho \omega_{zz} +\rho f_z \omega_z \omega_{\rho z} )\right. 
\nonumber\\
&& \left. +4 \rho \omega_\rho (-\rho \omega_\rho f_z \omega_{\rho z} + \rho \omega_\rho f_\rho \omega_{zz} + \omega_\rho^2 f_\rho + \rho \omega_\rho^2 f_{zz})
\right]f^6
\nonumber\\
&& + \left[-6 \rho^2 \omega_\rho^3 (f_z^2 + f_\rho^2) -2 \rho^2 f_z \omega_\rho \omega_z(2\omega_\rho f_\rho + 5 \omega_z f_z) \right.
\nonumber\\
&& \left. + 2 \rho^2 f_\rho \omega_z^2 (2 \omega_z f_z - \omega_\rho f_\rho) \right]f^5 + \left[8\rho^2(f_{\rho z} \omega_z - f_\rho \omega_{zz})\right.
\nonumber\\
&& \left. -16\rho^3(f_{\rho z} \omega_{\rho z} + f_{zz} \omega_{zz}) + 10 \rho^3 f_z \omega_z f_\rho \omega_\rho (f_z \omega_z + f_\rho \omega_\rho)\right.
\nonumber\\
&& \left. - 2\rho^3 f_z(f_\rho f_z \omega_\rho^3 + f_\rho^2 \omega_z^3) +2 \rho^3 f_\rho^3(\omega_\rho^3 - \omega_\rho \omega_z^2)\right.
\nonumber\\
&& \left. +2 \rho^3 f_z^3 (\omega_z^3 -\omega_z \omega_\rho^2)
\right] f^4 + \left[-24\rho^3 f_{\rho z}( \omega_\rho f_z + \omega_z f_\rho)\right.
\nonumber\\
&& \left. + 4 \rho^2 f_\rho (f_\rho \omega_\rho -4 f_z \omega_z) +4 \rho^3 (3 f_\rho^2 \omega_{zz} +2 f_{zz} f_\rho \omega_\rho -2 \omega_{\rho z} f_\rho f_z\right.
\nonumber\\ 
&&\left. + f_z^2 \omega_{zz} -10 f_{zz} \omega_z f_z)
\right] f^3 + \left[4 \rho^4 f_{\rho z}( f_\rho^2 \omega_z -f_z^2 \omega_z +2 \omega_\rho f_\rho f_z) \right.
\nonumber\\
&&\left. + 4 \rho^4f_{zz} (f_z^2 \omega_\rho - f_\rho^2 \omega_\rho + 2 \omega_z f_\rho f_z ) + 4 \rho^4 \omega_{\rho z} (-f_z^3 + 3f_z f_\rho^2)\right.
\nonumber\\
&& \left. +4 \rho^4 \omega_{zz} (-f_\rho^3 + 3 f_\rho f_z^2) + 4 \rho^3 (4 f_\rho^2 f_z \omega_z - 3 f_z^2 f_\rho \omega_\rho \right.
\nonumber\\ 
&& \left. - 2 f_\rho^3 \omega_\rho + 3 f_z^3 \omega_z)
\right] f^2 + \left[\rho^4 \omega_\rho (14 f_z^2 f_\rho^2 - 7f_z^4 +5f_\rho^4) \right.
\nonumber\\
&& \left.+ 4 \rho^4 \omega_z(f_\rho^3 f_z + 5 f_z^3 f_\rho)\right]f -\rho^5 \omega_z(2f_z^3 f_\rho^2+ f_z^5 + f_z f_\rho^4 )
\nonumber\\
&& - \rho^5 \omega_\rho (2 f_\rho^3 f_z^2 + f_\rho^5 + f_rho f_z^4 ).
\nonumber
\end{eqnarray}

\end{document}